\documentclass{book}    
\usepackage{piers}  
\begin{document}

\title{Mass continuity equation in the electromagnetic field}
\maketitle

\author      {Ying Weng}
\affiliation {Xiamen University}
\address     {}
\city        {Xiamen}
\postalcode  {}
\country     {China}
\phone       {345566}    
\fax         {233445}    
\email       {xmuwzh@hotmail.com}  
\misc        { }  
\nomakeauthor

\author      {Zi-Hua Weng}
\affiliation {Xiamen University}
\address     {}
\city        {Xiamen}
\postalcode  {}
\country     {China}
\phone       {345566}    
\fax         {233445}    
\email       {xmuwzh@xmu.edu.cn}  
\misc        { }  
\nomakeauthor


\begin{authors}

{\bf Ying Weng}$^{1}$, {\bf Zi-Hua Weng}$^{2}$ \\
\medskip
$^{1}$ College of Chemistry \& Chemical Engineering, \\Xiamen
University, Xiamen 361005, China\\
$^{2}$ School of Physics and Mechanical \& Electrical Engineering,
\\Xiamen University, Xiamen 361005, China\\

\end{authors}

\begin{paper}

\begin{piersabstract}
A theoretical method with the quaternion algebra was presented to
derive the mass continuity equation from the linear momentum. It
predicts that the strength of electromagnetic field and the velocity
have the impact on the mass continuity equation. In the
gravitational field and electromagnetic field, the mass continuity
equation will change with the electromagnetic field strength,
gravitational field strength, linear momentum, electric current, and
the speed of light. The deduction can explain why the field strength
has an influence on the anomalous transport about the mass
continuity equation in the plasma and electrolytes etc.
\end{piersabstract}


\psection{Introduction}

The conservation laws are important invariants for the
electromagnetic field. In the electromagnetic field theory described
by the vectorial quantity, the mass continuity equation is solely
dealt with the mass rate and the divergence of linear momentum.
However, this opinion can not explain why the field strength has an
impact on the anomalous transport about the mass continuity equation
in the plasma and electrolytes \cite{escande, ciraolo}.

The algebra of quaternion \cite{schwartz} was first used by J. C.
Maxwell \cite{maxwell} to describe the property of electromagnetic
field. Similarly, the quaternion can also be used to demonstrate the
feature of gravitational field, although these two fields are quite
different \cite{weng}. By means of the scalar invariant of
quaternions, we find that the mass continuity equation is an
invariant in the gravitational field and electromagnetic field. In
the quaternion spaces, the definition of mass continuity equation
can be extended to the case for coexistence of electromagnetic field
and gravitational field.

Presently, the mass continuity equation \cite{lavoisier} was limited
to the case of weak gravitational strength. It is found out that all
of related verifications are solely constrained to be in the range
of weak field strengths, and have not been validated in the strong
fields up to now. With the characteristics of octonions
\cite{cayley}, we find some impact factors of the mass continuity
equation. In the electromagnetic field and gravitational field, the
velocity and the strengths of electromagnetic field and
gravitational field have a few influences on the mass continuity
equation.

The results state that the gravitational field strength and
electromagnetic field strength have an influence on the mass
continuity equation, although the impact of the field strengths are
usually very tiny when the electromagnetic field and gravitational
field both are weak. And then, the mass continuity equation is
conserved in most cases. However, when the electromagnetic field and
gravitational field are strong enough, their field strengths will
affect the mass continuity equation obviously, and cause the
anomalous transport in the plasma.

\psection{Coordinates transformation}

In the quaternion space, the basis vector for the gravitational
field is $\mathbb{E}_g$ = ($1$, $\emph{\textbf{i}}_1$,
$\emph{\textbf{i}}_2$, $\emph{\textbf{i}}_3$), and that for the
electromagnetic field is $\mathbb{E}_e$ = ($\emph{\textbf{I}}_0$,
$\emph{\textbf{I}}_1$, $\emph{\textbf{I}}_2$,
$\emph{\textbf{I}}_3$). The $\mathbb{E}_e$ is independent of the
$\mathbb{E}_g$, with $\mathbb{E}_e$ = $\mathbb{E}_g \circ
\emph{\textbf{I}}_0$ . The basis vectors $\mathbb{E}_g$ and
$\mathbb{E}_e$ can be combined together to become the basis vector
$\mathbb{E}$ of the octonion space.
\begin{eqnarray}
\mathbb{E} = (1, \emph{\textbf{i}}_1, \emph{\textbf{i}}_2,
\emph{\textbf{i}}_3, \emph{\textbf{I}}_0, \emph{\textbf{I}}_1,
\emph{\textbf{I}}_2, \emph{\textbf{I}}_3)
\end{eqnarray}

The octonion quantity $\mathbb{D} (d_0, d_1, d_2, d_3, D_0, D_1,
D_2, D_3 )$ is defined as follows.
\begin{eqnarray}
\mathbb{D} = d_0 + \Sigma (d_j \emph{\textbf{i}}_j) + \Sigma (D_i
\emph{\textbf{I}}_i)
\end{eqnarray}
where, $d_i$ and $D_i$ are all real; $i = 0, 1, 2, 3$; $j = 1, 2,
3$.

When the coordinate system is transformed into the other one, the
physical quantity $\mathbb{D}$ will be become the octonion physical
quantity $\mathbb{D}' (d'_0 , d'_1 , d'_2 , d'_3 , D'_0 , D'_1 ,
D'_2 , D'_3 )$ .
\begin{equation}
\mathbb{D}' = \mathbb{K}^* \circ \mathbb{D} \circ \mathbb{K}
\end{equation}
where, $\mathbb{K}$ is the octonion, and $\mathbb{K}^* \circ
\mathbb{K} = 1$; $*$ denotes the conjugate of octonion; $\circ$ is
the octonion multiplication.

In case of the $d_0$ does not take part in the coordinates
transformation in the above, we have,
\begin{eqnarray}
d_0 = d'_0~, ~\mathbb{D}^* \circ \mathbb{D} = (\mathbb{D'})^* \circ
\mathbb{D'}~.
\end{eqnarray}

In the above equation, the scalar part is one and the same during
the octonion coordinates are transformed. Some invariants including
the mass continuity equation about the electromagnetic field will be
obtained from this characteristics of octonions.

\begin{table}[h]
\caption{The octonion multiplication table.} \label{tab:table1}
\centering
\begin{tabular}{ccccccccc}
\hline \hline $ $ & $1$ & $\emph{\textbf{i}}_1$  &
$\emph{\textbf{i}}_2$ & $\emph{\textbf{i}}_3$  &
$\emph{\textbf{I}}_0$  & $\emph{\textbf{I}}_1$
& $\emph{\textbf{I}}_2$  & $\emph{\textbf{I}}_3$  \\
\hline $1$ & $1$ & $\emph{\textbf{i}}_1$  & $\emph{\textbf{i}}_2$ &
$\emph{\textbf{i}}_3$  & $\emph{\textbf{I}}_0$  &
$\emph{\textbf{I}}_1$
& $\emph{\textbf{I}}_2$  & $\emph{\textbf{I}}_3$  \\
$\emph{\textbf{i}}_1$ & $\emph{\textbf{i}}_1$ & $-1$ &
$\emph{\textbf{i}}_3$  & $-\emph{\textbf{i}}_2$ &
$\emph{\textbf{I}}_1$
& $-\emph{\textbf{I}}_0$ & $-\emph{\textbf{I}}_3$ & $\emph{\textbf{I}}_2$  \\
$\emph{\textbf{i}}_2$ & $\emph{\textbf{i}}_2$ &
$-\emph{\textbf{i}}_3$ & $-1$ & $\emph{\textbf{i}}_1$  &
$\emph{\textbf{I}}_2$  & $\emph{\textbf{I}}_3$
& $-\emph{\textbf{I}}_0$ & $-\emph{\textbf{I}}_1$ \\
$\emph{\textbf{i}}_3$ & $\emph{\textbf{i}}_3$ &
$\emph{\textbf{i}}_2$ & $-\emph{\textbf{i}}_1$ & $-1$ &
$\emph{\textbf{I}}_3$  & $-\emph{\textbf{I}}_2$
& $\emph{\textbf{I}}_1$  & $-\emph{\textbf{I}}_0$ \\
\hline $\emph{\textbf{I}}_0$ & $\emph{\textbf{I}}_0$ &
$-\emph{\textbf{I}}_1$ & $-\emph{\textbf{I}}_2$ &
$-\emph{\textbf{I}}_3$ & $-1$ & $\emph{\textbf{i}}_1$
& $\emph{\textbf{i}}_2$  & $\emph{\textbf{i}}_3$  \\
$\emph{\textbf{I}}_1$ & $\emph{\textbf{I}}_1$ &
$\emph{\textbf{I}}_0$ & $-\emph{\textbf{I}}_3$ &
$\emph{\textbf{I}}_2$  & $-\emph{\textbf{i}}_1$
& $-1$ & $-\emph{\textbf{i}}_3$ & $\emph{\textbf{i}}_2$  \\
$\emph{\textbf{I}}_2$ & $\emph{\textbf{I}}_2$ &
$\emph{\textbf{I}}_3$ & $\emph{\textbf{I}}_0$  &
$-\emph{\textbf{I}}_1$ & $-\emph{\textbf{i}}_2$
& $\emph{\textbf{i}}_3$  & $-1$ & $-\emph{\textbf{i}}_1$ \\
$\emph{\textbf{I}}_3$ & $\emph{\textbf{I}}_3$ &
$-\emph{\textbf{I}}_2$ & $\emph{\textbf{I}}_1$  &
$\emph{\textbf{I}}_0$  & $-\emph{\textbf{i}}_3$
& $-\emph{\textbf{i}}_2$ & $\emph{\textbf{i}}_1$  & $-1$ \\
\hline
\end{tabular}
\end{table}

\psection{Velocity}

The radius vector is $\mathbb{R}_g$ = ($r_0$, $r_1$, $r_2$, $r_3$)
in the quaternion space for gravitational field. For the
electromagnetic field, the radius vector is $\mathbb{R}_e$ = ($R_0$,
$R_1$, $R_2$, $R_3$). Their combination is the radius vector
$\mathbb{R} (r_0 , r_1 , r_2 , r_3 , R_0 , R_1 , R_2 , R_3 )$ in the
octonion space.
\begin{eqnarray}
\mathbb{R} = r_0 + \Sigma (r_j \emph{\textbf{i}}_j) + \Sigma (R_i
\emph{\textbf{I}}_i)
\end{eqnarray}
where, $r_0 = v_0 t$. $v_0$ is the speed of light, $t$ denotes the
time.

In the quaternion space for the gravitational field, the velocity is
$\mathbb{V}_g$ = ($v_0$, $v_1$, $v_2$, $v_3$). For the
electromagnetic field, the velocity is $\mathbb{V}_e$ = ($V_0$,
$V_1$, $V_2$, $V_3$). They can be combined together to become the
octonion velocity $\mathbb{V} (v_0 , v_1 , v_2 , v_3 , V_0 , V_1 ,
V_2 , V_3 )$ in the octonion space.
\begin{eqnarray}
\mathbb{V} = v_0 + \Sigma (v_j \emph{\textbf{i}}_j) + \Sigma (V_i
\emph{\textbf{I}}_i)
\end{eqnarray}

In some cases, the electric charge is combined with the mass to
become the electron or proton etc., we have the relation $ R_i
\emph{\textbf{I}}_i = r_i \emph{\textbf{i}}_i \circ
\emph{\textbf{I}}_0$ and $V_i \emph{\textbf{I}}_i = v_i
\emph{\textbf{i}}_i \circ \emph{\textbf{I}}_0$ , with
$\emph{\textbf{i}}_0 = 1$.

\psection{Field Strength}

From the Aharonov-Bohm effect, we find that the field potential is
much more essential than the field strength. By means of the field
potential and the quaternion operator, the field strength can be
defined to cover the strengths of gravitational field and
electromagnetic field.

The gravitational potential is $\mathbb{A}_g = (a_0 , a_1 , a_2 ,
a_3)$, and the electromagnetic potential is $\mathbb{A}_e = (A_0 ,
A_1 , A_2 , A_3)$. The gravitational potential and electromagnetic
potential constitute the field potential $\mathbb{A}(a_0 , a_1 , a_2
, a_3 , A_0 , A_1 , A_2 , A_3 )$ in the octonion space.
\begin{eqnarray}
\mathbb{A} = \mathbb{A}_g + k_{eg} \mathbb{A}_e
\end{eqnarray}
where, $k_{eg}$ is the coefficient.

The octonion strength $\mathbb{B}$ consists of gravitational
strength $\mathbb{B}_g$ and electromagnetic strength $\mathbb{B}_e$.
\begin{eqnarray}
\mathbb{B} = \lozenge \circ \mathbb{A} = \mathbb{B}_g + k_{eg}
\mathbb{B}_e
\end{eqnarray}
where, the quaternion operator $\lozenge = \Sigma (
\emph{\textbf{i}}_i \partial_i )$ , with $\partial_i = \partial /
\partial r_i$ .

In the above equation, we choose two gauge conditions of field
potential to simplify succeeding calculation. The gravitational
potential gauge is $ b_0 = \partial_0 a_0 + \nabla \cdot \textbf{a}
= 0$ , and the electromagnetic potential gauge $B_0 = \partial_0 A_0
+ \nabla \cdot \textbf{A} = 0$. Where, $\textbf{a} = \Sigma (a_j
\emph{\textbf{i}}_j);~ \textbf{A} = \Sigma (A_j
\emph{\textbf{i}}_j);~\nabla = \Sigma (\emph{\textbf{i}}_j
\partial_j)$.

The gravitational field strength $\mathbb{B}_g = \textbf{g}/c +
\textbf{b}$ includes two components, $\textbf{g}/c = \partial_0
\textbf{a} + \nabla a_0$ and $\textbf{b} = \nabla \times \textbf{a}$
. At the same time, the electromagnetic field strength $\mathbb{B}_e
= \textbf{E}/c + \textbf{B}$ involves two parts, $\textbf{E}/c = (
\partial_0 \textbf{A} + \nabla A_0 ) \circ \emph{\textbf{I}}_0 $ and
$\textbf{B} = -( \nabla \times \textbf{A} ) \circ
\emph{\textbf{I}}_0$ .

In the Newtonian gravitational field theory, there are $\textbf{a} =
0$ and $\textbf{b} = 0$ specially.

\psection{Linear momentum}

The linear momentum is one function of the field source, which is a
combination of the gravitational field source and electromagnetic
field source.

The linear momentum density $\mathbb{S}_g = m \mathbb{V}_g $ is the
source of the gravitational field, and the electric current density
$\mathbb{S}_e = q \mathbb{V}_g \circ \emph{\textbf{I}}_0$ is that of
the electromagnetic field. They combine together to become the field
source $\mathbb{S}$ .
\begin{eqnarray}
\mu \mathbb{S} = - ( \mathbb{B}/v_0 + \lozenge)^* \circ \mathbb{B} =
\mu_g \mathbb{S}_g + k_{eg} \mu_e \mathbb{S}_e - \mathbb{B}^* \circ
\mathbb{B}/v_0
\end{eqnarray}
where, $k_{eg}^2 = \mu_g /\mu_e$; $q$ is the electric charge
density; $m$ is the inertial mass density; $\mu$, $\mu_g$, and
$\mu_e$ are the constants.

The $\mathbb{B}^* \circ \mathbb{B}/(2\mu_g)$ is the energy density,
and includes that of the electromagnetic field.
\begin{eqnarray}
\mathbb{B}^* \circ \mathbb{B}/ \mu_g = \mathbb{B}_g^* \circ
\mathbb{B}_g / \mu_g + \mathbb{B}_e^* \circ \mathbb{B}_e / \mu_e
\end{eqnarray}

The octonion linear momentum density is
\begin{eqnarray}
\mathbb{P} = \mu \mathbb{S} / \mu_g = \widehat{m} v_0 + \Sigma (m
v_j \emph{\textbf{i}}_j ) + \Sigma (M V_i \emph{\textbf{i}}_i \circ
\emph{\textbf{I}}_0 )
\end{eqnarray}
where, $\widehat{m} = m + \triangle m $; $M = k_{eg} \mu_e q /
\mu_g$; $\triangle m = - ( \mathbb{B}^* \circ
\mathbb{B}/\mu_g)/v_0^2$ .

The above means that the gravitational mass density $\widehat{m}$ is
changed with the strength of either electromagnetic field or
gravitational field.

\psection{Mass continuity equation}

The applied force can not be covered by Maxwell's equations, which
are derived from the definition of field source. Whereas, the
applied force can be derived from the linear momentum. And that the
applied force covers the mass continuity equation in the
gravitational and electromagnetic fields.

In the octonion space, the applied force density $\mathbb{F}$ is
defined from the linear momentum density $\mathbb{P}$ in the
gravitational field and electromagnetic field.
\begin{eqnarray}
\mathbb{F} = v_0 (\mathbb{B}/v_0 + \lozenge )^* \circ \mathbb{P}
\end{eqnarray}
where, the applied force includes the gravity, the inertial force,
the Lorentz force, and the interacting force between the magnetic
strength with magnetic moment, etc.

The applied force density $\mathbb{F}$ is rewritten as follows.
\begin{eqnarray}
\mathbb{F} = f_0 + \Sigma (f_j \emph{\textbf{i}}_j ) + \Sigma (F_i
\emph{\textbf{I}}_i )
\end{eqnarray}
where, $f_0 = \partial p_0 / \partial t + v_0 \Sigma (  \partial p_j
/ \partial r_j ) + \Sigma ( b_j p_j + B_j P_j ) $; $p_0 =
\widehat{m} v_0$, $p_j = m v_j $; $P_i = M V_i $ .

We have the octonion applied force density $\mathbb{F}' (f'_0, f'_1,
f'_2, f'_3, F'_0, F'_1, F'_2, F'_3)$, when the coordinate system
rotates. And then, we have the following result by Eq.(4).
\begin{eqnarray}
f_0 = f'_0
\end{eqnarray}

When the right side is zero in the above, we have the mass
continuity equation in the case for coexistence of the gravitational
field and electromagnetic field.
\begin{eqnarray}
\partial \widehat{m} / \partial t + \Sigma (  \partial p_j /
\partial r_j ) + \Sigma ( b_j p_j  + B_j P_j ) / v_0 = 0
\end{eqnarray}

Further, if the strength is zero, $b_j = B_j = 0$, the above will be
reduced as follows.
\begin{eqnarray}
\partial m / \partial t + \Sigma (  \partial p_j /
\partial r_j ) = 0
\end{eqnarray}

The above states that the gravitational strength and electromagnetic
strength have the influence on the mass continuity equation,
although the $\Sigma ( b_j p_j +  B_j P_j ) / v_0$ and the
$\triangle m$ both are usually very tiny when the fields are weak.
When we emphasize the definitions of applied force and velocity in
gravitational and electromagnetic fields, the mass continuity
equation will be the invariant equation under the octonion
transformation.

\psection{Conclusion}

The mass continuity equation will vary in the strong electromagnetic
field or gravitational field, and has a deviation from the mass
continuity equation described with the vectorial quantity. In the
gravitational and electromagnetic fields, this states that the mass
continuity equation will change with the electromagnetic field
strength, gravitational field strength, linear momentum, electric
current, and the speed of light. The deduction can explain why the
field strength has an impact on the anomalous transport about the
mass continuity equation in the plasma.

In the octonion space, the deductive results about the conservation
laws and the invariants depend on the definition combinations in the
case for coexistence of gravitational field and electromagnetic
field. By means of the definition combination of the linear
momentum, the quaternion operator, and the velocity, we have the
conclusions about the mass continuity equation in the gravitational
field and electromagnetic field.

It should be noted that the study for the mass continuity equation
examined only one simple case of weak field strength. Despite its
preliminary characteristics, this study can clearly indicate the
mass continuity equation is an invariant and is only one simple
inference due to the low velocity and the weak strengths of
electromagnetic field and gravitational field. For the future
studies, the investigation will concentrate on only some predictions
about the mass continuity equation under the high speed and strong
strength of electromagnetic field.

\ack
This project was supported partially by the National Natural
Science Foundation of China under grant number 60677039.

\end{paper}

\end{document}